\documentstyle[aps, preprint, epsf]{revtex}
\begin{document}

\title{Signature of a Pairing Transition in the Heat Capacity
 of Finite Nuclei}

\author{S. Liu and Y. Alhassid }

\address{Center for Theoretical Physics, Sloane Physics Laboratory,
Yale University, New Haven, Connecticut  06520, U.S.A. }
\date{\today}
\maketitle

\begin{abstract}

  The heat capacity of iron isotopes is calculated within the
 interacting shell model using the complete $(pf+0g_{9/2})$-shell. 
We identify a signature of the pairing transition in the heat 
capacity that is correlated with the suppression of the number 
of spin-zero neutron pairs as the  temperature increases. 
Our results are obtained by a novel method that significantly
 reduces the statistical errors in the heat capacity calculated
 by the shell model Monte Carlo approach. The Monte Carlo results 
are compared with finite-temperature Fermi gas and BCS calculations.

\end{abstract}


Pairing effects in finite nuclei are well known; examples include 
the energy gap in the spectra of even-even nuclei and an odd-even
 effect observed in nuclear masses. However, less is known about
 the thermal signatures of the pairing interaction in nuclei.
 In a macroscopic conductor, pairing leads to a phase transition 
from a normal metal to a superconductor below a certain critical 
temperature, and in the BCS theory \cite{BCS} the heat capacity
 is characterized by a finite discontinuity at the transition
 temperature. As the linear dimension of the system decreases
 below the pair coherence length, fluctuations in the order
 parameter become important and lead to a smooth transition.
 The effects of both static fluctuations \cite{SPA72,SPA}
 and small quantal fluctuations \cite{Lauritzen93} have 
been explored in studies of small metallic grains. A pronounced
 peak in the heat capacity is observed for a large number of 
electrons, but for less than $\sim 100$ electrons the peak 
in the heat capacity all but disappears.  In the nucleus, 
the pair coherence length is always much larger than the 
nuclear radius,  and large fluctuations are expected to 
suppress any singularity in the heat capacity.  An interesting 
question is whether any signature of the pairing transition
 still exists in the heat capacity of the nucleus despite 
the large fluctuations. 
When only static and small-amplitude quantal fluctuations 
are taken into account, a shallow `kink' could still be seen in
 the heat capacity of  an even-even nucleus \cite{RCP98}. This
 calculation, however, was limited to a schematic pairing model. 
Canonical heat capacities were recently extracted from level density
 measurements in rare-earth nuclei \cite{s-shape} and were found
 to have an $S$-shape that is interpreted to represent the 
suppression of pairing correlations with increasing temperature.

The calculation of the heat capacity of the finite interacting
 nuclear system beyond the mean-field and static-path 
approximations is a difficult problem. Correlation effects 
due to residual interactions can be accounted for in the
 framework of the interacting nuclear shell model. However,
 at finite temperature a large number of excited states contribute
 to the heat capacity and  very large model spaces are necessary
 to obtain reliable results. The shell model Monte Carlo (SMMC) 
method \cite{SMMC,ADK94} enables zero- and finite-temperature
 calculations in large spaces. In particular, the thermal energy
 $E(T)$ can be computed versus temperature $T$ and the heat 
capacity can be obtained by taking a numerical derivative 
$C=dE/dT$. However, the finite statistical errors in $E(T)$ 
lead to large statistical errors in the heat capacity at low 
temperatures (even for good-sign interactions). 
Such large errors occur already around the pairing transition 
temperature and thus no definite signatures of the pairing 
transition could be identified. Furthermore, the large errors 
often lead to spurious structure in the calculated heat capacity.
  Presumably, a more accurate heat capacity can be obtained by 
a direct calculation of the variance of the Hamiltonian, but in
 SMMC such a calculation is impractical since it involves a 
four-body operator. The variance of the Hamiltonian has been 
calculated using a different Monte Carlo algorithm \cite{RHJ98},
 but that method is presently limited to a schematic pairing 
interaction.   Here we report a novel method for calculating 
the heat capacity within SMMC that takes into account correlated
 errors and leads to much smaller statistical errors. Using 
this method we are able to identify a signature of the pairing 
transition in {\em realistic} calculations of the heat capacity 
of finite nuclei. The signature is well correlated with the 
suppression in the number of spin-zero pairs across the transition
 temperature. 

The Monte Carlo approach is based on the Hubbard-Stratonovich 
(HS) representation of the many-body
imaginary-time propagator, $e^{-\beta  H}= \int D[\sigma] 
G_\sigma U_\sigma$, where $\beta$ is the inverse temperature, 
$G_\sigma$ is a Gaussian weight and $U_\sigma$ is a one-body 
propagator that describes non-interacting nucleons moving in
fluctuating time-dependent auxiliary fields $\sigma$.  
  The canonical
thermal expectation value of an observable $O$ can be written 
as $\langle O
\rangle= \int D[\sigma] G_\sigma{\rm Tr}\,(OU_\sigma)/\int
D[\sigma]  G_\sigma{\rm Tr}\, U_\sigma$, where ${\rm Tr}$
denotes a canonical trace for $N$ neutrons and $Z$ protons. 
 We can rewrite
\begin{equation}\label{observable}
\langle O \rangle = {\left\langle \left[{{\rm Tr}\,(O U_\sigma) /
{\rm  Tr}\, U_\sigma} \right] \Phi_\sigma\right\rangle_W \over
 \langle \Phi_\sigma \rangle_W} \;,
\end{equation}
where $\Phi_\sigma = {\rm  Tr}\, U_\sigma/| {\rm  Tr}\, U_\sigma|$ 
is the Monte Carlo sign, and we have used the notation
$\langle X_\sigma \rangle_W \equiv \int D[\sigma] W(\sigma) 
X_\sigma / \int  D[\sigma] W(\sigma)$
with $W(\sigma)\equiv G_\sigma |{\rm Tr} \, U_\sigma|$. 
In SMMC we divide the imaginary-time interval  $(0,\beta)$ 
 into $N_t$ time slices of length $\Delta \beta = \beta/N_t$,
 and sample the fields $\sigma(\tau_n)$ fields at all $N_t$ time 
slices $\tau_n = n\Delta \beta$ according to the positive-definite
 weight function $W(\sigma)$. Each quantity 
$\langle X_\sigma \rangle_W$ in Eq. (\ref{observable}) is then 
estimated as an arithmetic average over the chosen samples. 

   In particular the thermal energy can be calculated as a 
thermal average of the Hamiltonian $H$. The heat capacity 
$C = -\beta^2 \partial E /\partial \beta$ is then calculated 
by estimating the derivative as a finite difference
\begin{eqnarray} \label{deriv}
  C =
  -\beta^2\frac{E(\beta +\delta \beta ) -
    E(\beta -\delta \beta )}{2\delta \beta } + O(\delta \beta )^{2} \;.
\end{eqnarray}

At low temperatures, $E(\beta)$ decreases slowly with $\beta$ 
and even small errors in $E(\beta)$ lead to relatively large
 statistical errors in $C$. Conventionally, the calculation 
of $E(\beta)$ for each $\beta$ is done by a new Monte Carlo 
sampling of the $\sigma$ fields and consequently the energies
 $E(\beta - \delta \beta )$  and $E(\beta + \delta \beta )$ in
 (\ref{deriv}) are uncorrelated. However, if the calculation of
 both $E(\beta \pm \delta \beta)$ can be done using a common set
 of sampling fields, the correlated errors of $C$ are expected
 to be smaller. 

The energies $E(\beta \pm \delta\beta)$ are calculated from
\begin{equation}
  \label{orig-energy}
  E(\beta \pm \delta \beta) =
  {\int D[\sigma^\pm ] G_{\sigma^\pm}(\beta \pm \delta\beta)
    {\rm Tr}\,[HU_{\sigma^\pm}(\beta \pm \delta \beta)] \over
    \int D[\sigma^\pm ]G_{\sigma^\pm}(\beta \pm \delta\beta)
    {\rm Tr}\,U_{\sigma^\pm}(\beta \pm \delta \beta)} \;,
\end{equation}
where the corresponding $\sigma$ fields are denoted by 
$\sigma^\pm$. To have the same number of time slices $N_t$ 
 in the discretized version of (\ref{orig-energy}) as in the 
original HS representation of $E(\beta)$, we define modified
 time slices  $\Delta\beta_\pm$ by 
$N_t \Delta\beta_\pm = \beta \pm \delta\beta$. We next change 
integration variables in (\ref{orig-energy}) from $\sigma^\pm$
 to $\sigma$  according to $\sigma^\pm = 
(\Delta \beta/\Delta \beta_\pm)^{1/2} \sigma$, so that the
 Gaussian weight is left unchanged
 $G_{\sigma^\pm}(\beta \pm \delta\beta)
 \equiv \exp\left[-\sum\limits_{\alpha n} {1 \over 2}
  \vert v_\alpha \vert (\sigma^\pm_\alpha(\tau_n))^2 
\Delta\beta_\pm \right] = \exp\left[-\sum\limits_{\alpha n} 
{1 \over 2}
  \vert v_\alpha \vert (\sigma_\alpha(\tau_n))^2 
\Delta\beta \right] = G_\sigma(\beta)$ ($v_\alpha$ are 
the interaction  `eigenvalues', obtained by writing the
 interaction in a quadratic form $\sum_\alpha v_\alpha 
\hat \rho_\alpha^2/2$, where $\hat \rho_\alpha$ are 
one-body densities). 
Rewriting (\ref{orig-energy}) using the measure $D[\sigma]$ 
(the Jacobian resulting from the change in integration variables
 is constant and canceled between the numerator and denominator),
 we find 
\begin{equation}
  \label{energy}
  E(\beta \pm \delta \beta)
  = \frac{\left \langle
      \frac{{\rm Tr}\,HU_{\sigma^\pm}(\beta \pm \delta \beta )}
      {{\rm Tr}\, U_{\sigma^\pm}(\beta \pm \delta \beta)}
      \frac{{\rm Tr}\, U_{\sigma^\pm}(\beta \pm \delta \beta )}
      {{\rm Tr}\, U_\sigma(\beta)}
      \Phi_\sigma \right\rangle_W}
  {\left \langle
      \frac{{\rm Tr}\,U_{\sigma^\pm}(\beta \pm \delta \beta )}
      {{\rm Tr}\,U_\sigma(\beta)}
      \Phi_\sigma \right\rangle_W} \equiv {H_\pm \over Z_\pm} \;.
\end{equation}
  The heat capacity in (\ref{deriv}) is calculated from $C = -\beta^2 (2\delta\beta)^{-1}(H_+/Z_+ -H_-/Z_-)$.
Since the same set of fields $\sigma$ is used in the calculation
 of both $E(\beta\pm \delta\beta)$, we expect strong correlations
 among the quantities $H_\pm$ and $Z_\pm$, which would lead to
 a smaller error for $C$. 
 The covariances 
among $H_\pm$ and $Z_\pm$ as well as their variances can be 
calculated in the Monte Carlo and used to estimate the correlated
 error of the heat capacity.

\begin{figure}

\vspace{5 mm}

\centerline{\epsffile{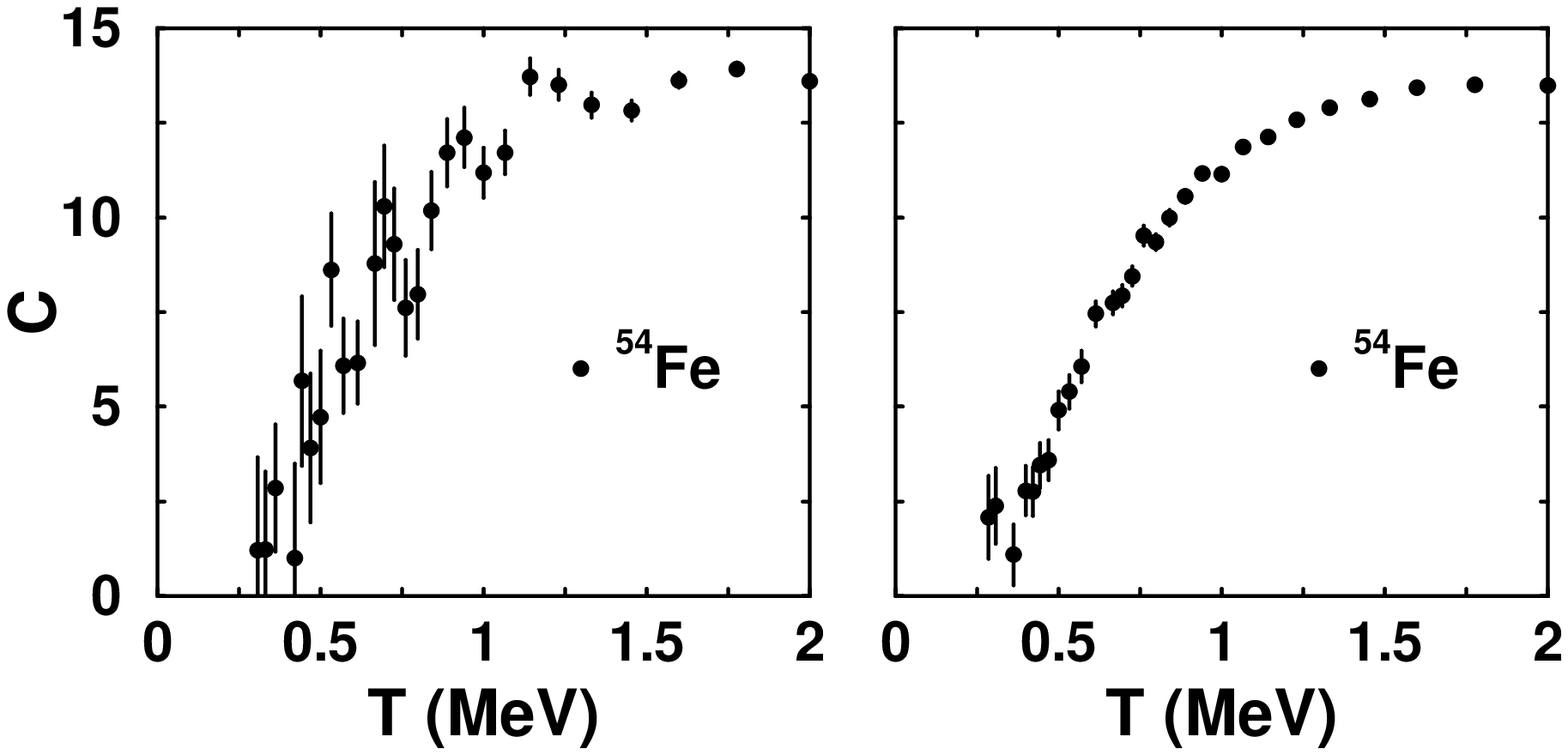}}

\vspace{3 mm}

 \caption { The SMMC heat capacity of $^{54}$Fe. The left panel
 is the result of conventional SMMC calculations. The right panel
 is calculated using the improved method (based on the representation (\protect\ref{energy}) where a correlated error can be accounted for).}
  \label{fig1}

\newpage

\centerline{\epsffile{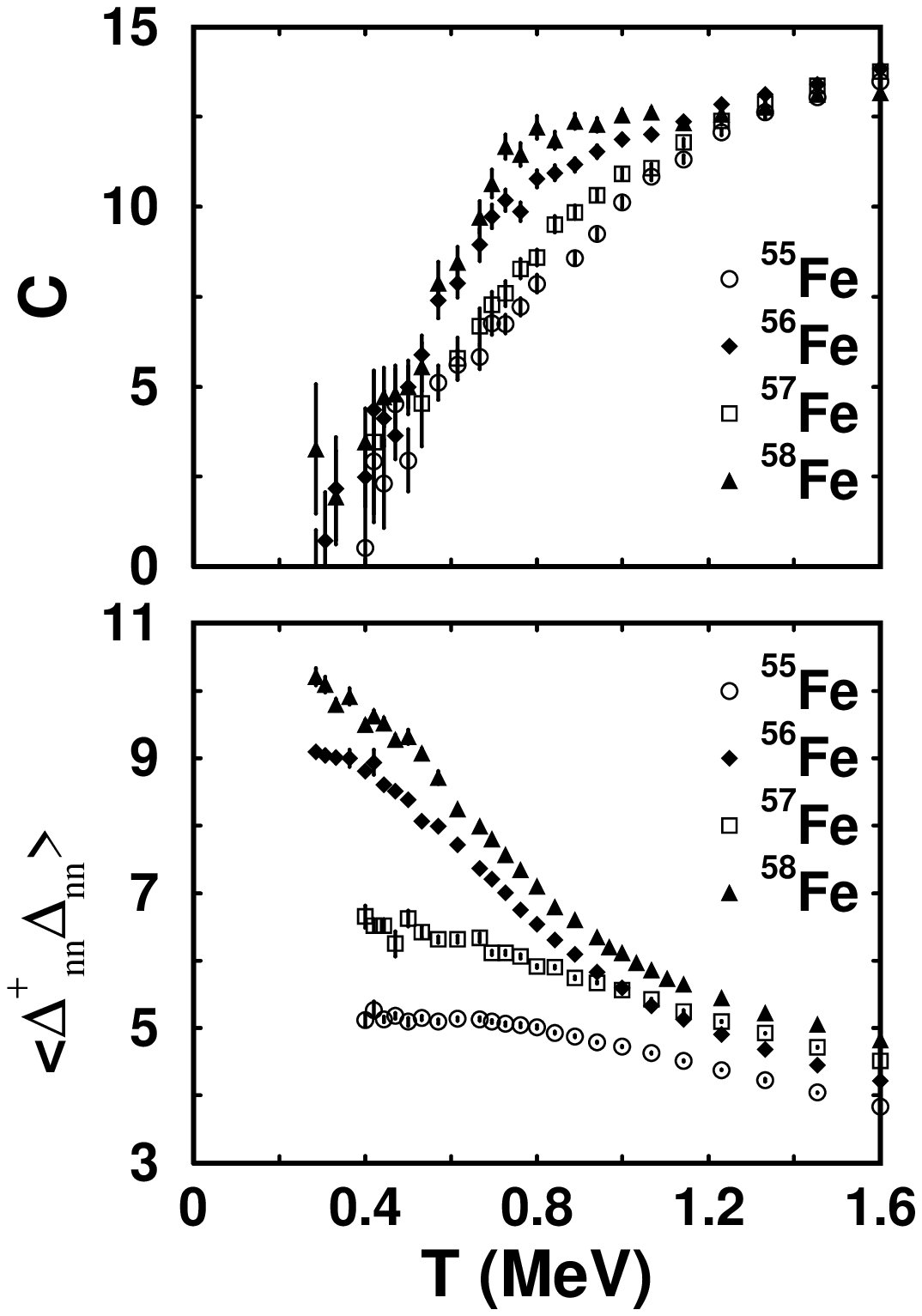}}

\vspace{3 mm}

  \caption { Top panel: the SMMC heat capacity vs. temperature
 $T$ for $^{55}$Fe (open circles),  $^{56}$Fe (solid diamonds),
 $^{57}$Fe (open squares), 
    and $^{58}$Fe (solid triangles). Bottom
    panel: the number of $J=0$ neutron pairs versus temperature
 for the same nuclei. }
  \label{fig2}

\newpage

 \centerline{\epsffile{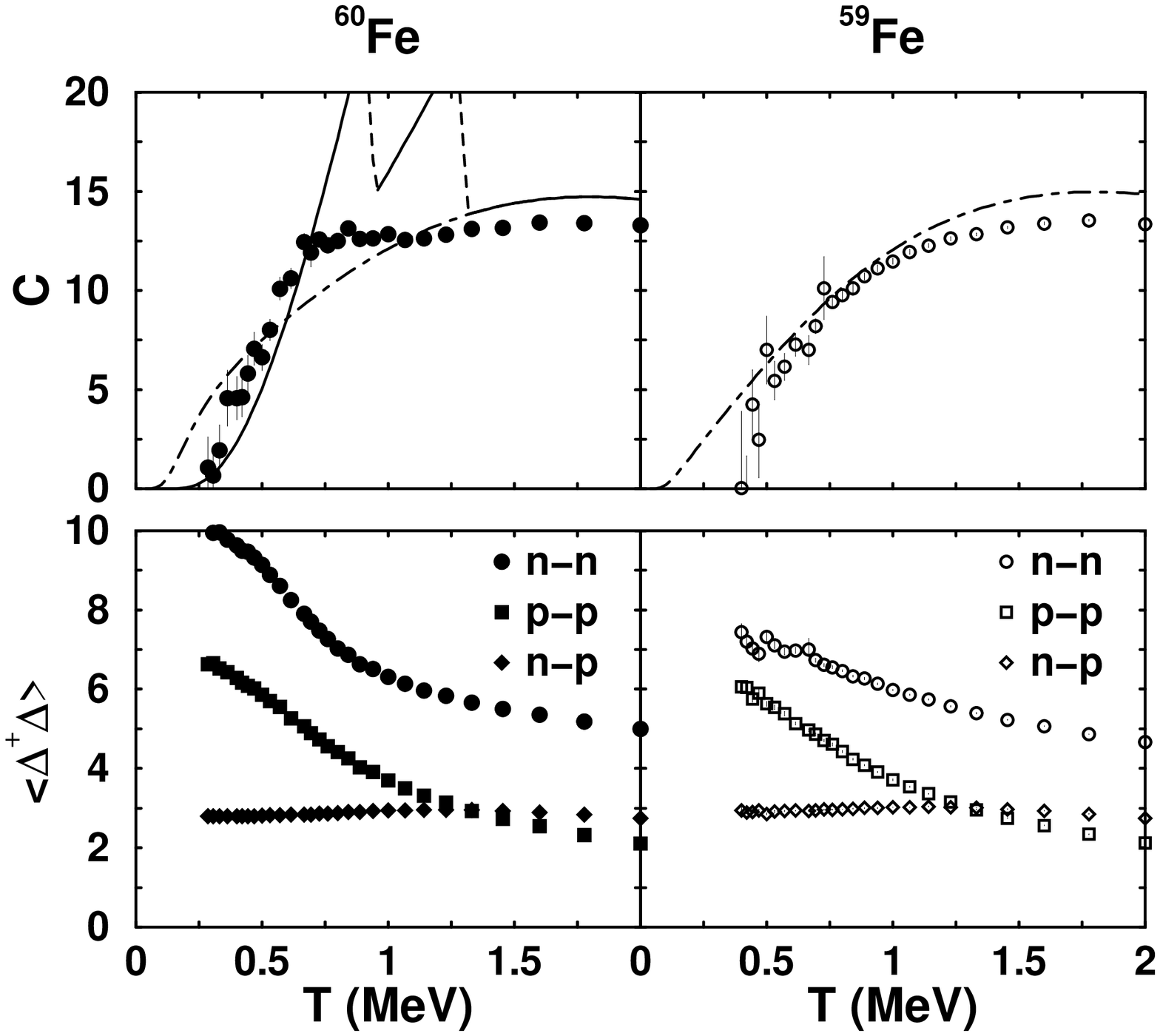}}

\vspace{3 mm}

  \caption { Top: Heat capacity  versus $T$  for $^{60}$Fe 
(left) and $^{59}$Fe (right). The Monte Carlo results are shown
 by symbols.  The dotted-dashed lines are the Fermi gas calculations,
 and the solid line (left panel only) is the BCS result. The
 discontinuities (dashed lines) correspond to a neutron ($T_c \sim 0.9$ MeV)
 and proton ($T_c \sim 1.2$ MeV) pairing transition.  Above the
 pairing-transition temperature, the BCS results coincide with 
the Fermi gas results.   Bottom panels: The number of $J=0$
 $n$-$n$ (circles),  $p$-$p$(squares), and $n$-$p$(diamonds)
 pairs vs. $T$ for $^{60}$Fe (left) and $^{59}$Fe (right).}
  \label{fig3}

\newpage

\centerline{\epsffile{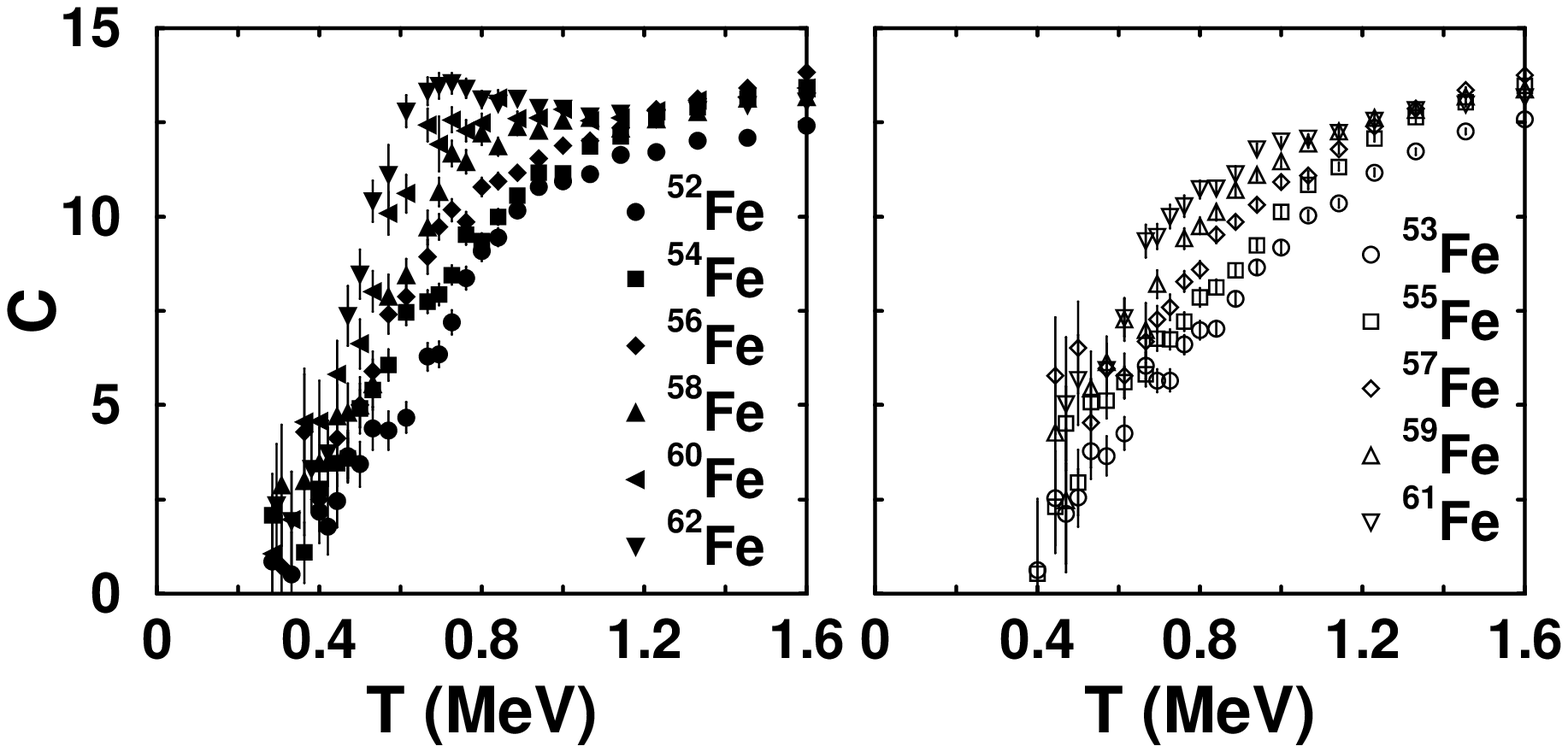}}
  \caption {The heat capacity of even-even (left panel) and 
odd-even (right panel) iron isotopes.}
  \label{fig4}
\end{figure}

We have calculated the heat capacity for the iron isotopes
$^{52-62}$Fe using the complete $(pf+0g_{9/2})$-shell and the
good-sign interaction of Ref. \cite{NA97}. Fig. \ref{fig1}
 demonstrates the 
significant improvement in the statistical Monte Carlo errors.
 On the left panel of this figure we 
show the heat capacity of $^{54}$Fe 
 calculated in the conventional method, while the right panel 
shows the results from the new method. The statistical errors 
for $T \sim 0.5 -1$ MeV are reduced by almost an order of 
magnitude. The results obtained in the conventional calculation 
seem to indicate a shallow peak in the heat capacity around 
$T \sim 1.25$ MeV, but the calculation using the improved method
shows no such structure. 

The heat capacities of four iron isotopes $^{55-58}$Fe, 
calculated with the new method, are shown in the top panel 
of Fig. \ref{fig2}. The heat capacities of the two even-mass 
iron isotopes ($^{56}$Fe and $^{58}$Fe) show a different 
behavior around $T \sim 0.7 -0.8$ MeV as compared with the 
two odd-mass isotopes ($^{55}$Fe and $^{57}$Fe). While the heat
 capacity of the odd-mass isotopes  
increases smoothly as a function of temperature, the heat capacity
 of the even-mass isotopes is enhanced for $T \sim 0.6 -1$ MeV,
 increasing  sharply and then leveling off, displaying a
 `shoulder.' This `shoulder'  is more pronounced for the isotope
 with more neutrons ($^{58}$Fe). 
To correlate this behavior of the heat capacity with a pairing
 transition, we calculated the number of $J=0$ nucleon pairs 
in these nuclei. A $J=0$ pair operator is defined as usual by  
\begin{equation}
  \label{pair}
  \Delta^\dagger =
  \sum\limits_{a, m_a >0}
  \frac{(-1)^{j_a - m_a}}{\sqrt{j_a + 1/2}}
  a^{\dagger}_{j_a m_a}a^{\dagger}_{j_a -m_a} \;,
\end{equation}
where $j_a$ is the spin and $m_a$ is the spin projection of 
a single-particle orbit $a$.  
Pair-creation operators of the form (\ref{pair}) can be
 defined for protons ($\Delta^\dagger_{pp}$), neutrons 
($\Delta^\dagger_{nn}$), and proton-neutrons 
($\Delta^\dagger_{pn}$). The average number
 $\langle \Delta^\dagger \Delta \rangle$ of $J=0$ pairs  
(of each type) can be calculated exactly in SMMC as a 
function of temperature. 
 The bottom panel of Fig. \ref{fig2} shows the number of
 neutron pairs $\langle \Delta^\dagger_{nn} \Delta_{nn}\rangle$
 for $^{55-58}$Fe. 
At low temperature the number of neutron pairs for isotopes
 with an even number of neutrons is 
significantly larger than that for isotopes with an odd
 number of neutrons. Furthermore, for the even-mass isotopes
 we observe a rapid suppression of the number of neutron
 pairs that correlates with the `shoulder' observed in the 
heat capacity. The different  qualitative behavior in the 
number of neutron pairs versus temperature between
odd- and even-mass iron isotopes provides a clue to the
 difference in their heat capacities. A transition from a 
pair-correlated ground state to a normal state at higher 
temperatures requires additional energy for breaking of 
neutron pairs, hence the steeper increase observed in 
the heat capacity of the even-mass iron isotopes. Once 
the pairs are broken, less energy is required to increase
 the temperature, and the heat capacity shows only
 a moderate increase. 

 It is instructive to compare the SMMC heat capacity with 
a Fermi gas and BCS calculations. The heat capacity can 
be calculated from the entropy using the relation 
  $C = T \partial S / \partial T$. The entropy $S$ of
 uncorrelated fermions is given by
\begin{equation}
  \label{entropy}
  S(T) = - \sum\limits_a [ f_a \ln f_a + (1-f_a) \ln(1-f_a) ] \;,
\end{equation}
with $f_a$ being the finite-temperature occupation numbers
 of the single-particle orbits $a$. Above the pairing
 transition-temperature $T_c$, $f_a$ are just the Fermi-Dirac 
occupancies
$f_a = [1 + e^{\beta(\epsilon_a - \mu)}]^{-1}$,
where $\mu$ is the chemical potential determined from 
the total number of particles and $\epsilon_a$ are the 
single-particle energies. Below $T_c$, it is necessary to
 take into account the BCS solution which has lower free energy.
  Since condensed pairs do not contribute to the entropy, the
 latter is still given by (\ref{entropy}) but $f_a$ are now the
 quasi-particle occupancies  \cite{BCS},
\begin{equation}
  \label{quasi}
  f_a = \frac{1}{ 1 + e^{\beta E_a} } \;.
\end{equation}
$E_a = \sqrt{(\epsilon_a -\mu)^2 + \Delta^2}$ are the quasi-particle 
energies, where the gap $\Delta(T)$ and the chemical potential
 $\mu(T)$ are determined from  
the finite-temperature BCS equations. In practice, we
treat protons and neutrons separately. 

We applied the Fermi gas and BCS approximations to estimate 
the heat capacities of the iron isotopes. To take into account
 effects of a quadrupole-quadrupole interaction, we used an
 axially deformed 
Woods-Saxon potential to extract the single-particle spectrum 
$\epsilon_a$ \cite{parity}.  A deformation parameter $\delta$
 for the even iron isotopes can be extracted from experimental 
$B(E2)$ values. However, since $B(E2)$ values are not available
 for all of these isotopes, we  used an alternate procedure.
  The excitation energy $E_x(2_1^+)$ of the first excited 
$2^+$ state in even-even nuclei can be extracted in SMMC
 by calculating $\langle J^2 \rangle_\beta$ at low temperatures
 and using a two-state model (the $0^+$ ground state and 
the first excited $2^+$ state) where $\langle J^2 
\rangle_\beta \approx 6 / ( 1+e^{\beta E_x(2_1^+)}/5)$ \cite{NA98}.
  The excitation energy of the $2_1^+$ state is then used 
in the empirical formula of Bohr and Mottelson \cite{raman}
$\tau_\gamma = (5.94\pm 2.43)\times 10^{14} E_x^{-4}(2_1^+)
 Z^{-2} A^{1/3}$ to estimate the  
mean $\gamma$-ray lifetime $\tau_\gamma$ and the corresponding
 $B(E2)$.  The deformation parameter $\delta$ is then estimated
 from  $B(E2) = [(3/4\pi)Zer_0^2 A^{2/3} \delta]^2 / 5$.
We find (using $r_0 = 1.27$ fm) $\delta = 0.225, 0.215, 0.244,
 0.222, 0.230$,
and $0.220$ for the even iron isotopes $^{52}$Fe -- $^{62}$Fe,
respectively. For the odd-mass iron isotopes we adapt the deformations
in Ref. \cite{DEF}. The zero-temperature pairing gap 
$\Delta$ is extracted from experimental odd-even mass 
differences and used to determine 
the pairing strength $G$ (needed for the finite temperature
 BCS solution).

The top panels of Fig. \ref{fig3} show the Fermi-gas heat 
capacity (dotted-dashed lines) for $^{59}$Fe (right) and 
$^{60}$Fe (left) in comparison with the SMMC results (symbols).
 The SMMC heat capacity in the even-mass $^{60}$Fe is below
 the Fermi-gas estimate for $T \alt 0.5$ MeV, but is enhanced 
above the Fermi gas heat capacity in the region $0.5 
\alt T\alt 0.9$ MeV.  The line shape of the heat capacity is 
similar to the $S$-shape found experimentally in the heat 
capacity of rare-earth nuclei \cite{s-shape}. We remark 
that the saturation of the SMMC heat capacity above 
$\sim 1.5 $ MeV (and eventually its decrease with $T$)
 is an artifact of the finite model space. The solid 
line shown for $^{60}$Fe is the result of the BCS calculation.
 There are two `peaks' in the heat capacity corresponding 
to separate pairing transitions for neutrons 
($T^n_c \approx 0.9$ MeV) and protons ($T^p_c \approx 1.2$ MeV). 
The finite discontinuities in the BCS heat capacity are shown
 by the dotted lines. The pairing solution describes well 
the SMMC results for $T \alt 0.6$ MeV. However, the BCS peak
 in the heat capacity is strongly suppressed around the 
transition temperature. This is expected in the finite nuclear
 system because of the strong fluctuations in the vicinity 
of the pairing transition (not accounted for in the mean-field 
approach). Despite the large fluctuations, a `shoulder' still
 remains around the neutron-pairing transition temperature. 

The bottom panels of Fig. \ref{fig3} show the number of 
spin-zero pairs versus temperature in SMMC. The number 
of $p$-$p$ and $n$-$p$ pairs are similar in the even and 
odd-mass iron isotopes. However, the number of $n$-$n$ pairs 
at low $T$ differs significantly between the two isotopes. 
The $n$-$n$ pair number of $^{60}$Fe decreases rapidly as 
a function of $T$, while that of $^{59}$Fe decreases slowly. 
The S-shape or shoulder seen
in the SMMC heat capacity of $^{60}$Fe correlates well with 
 the suppression of neutron pairs. 

Fig. \ref{fig4} shows the complete systematics of the heat
 capacity for the
iron isotopes in the mass range $A =52- 62$ for both even-mass 
(left panel) and odd-mass (right panel).
At low temperatures the heat capacity approaches zero, as expected.
When $T$ is high, the heat capacity for all isotopes converges to 
approximately the same value. In the intermediate temperature
 region ($T \sim 0.7$ MeV), the heat capacity increases with 
mass due to the increase of the density of states with mass.
 Pairing leads to an odd-even staggering effect in the 
mass dependence (see also in Fig. \ref{fig2}) where the heat 
capacity of an odd-mass nucleus is significantly 
lower than that of the adjacent even-mass nuclei. For example,
 the heat capacity of $^{57}$Fe is below that of 
both $^{56}$Fe and $^{58}$Fe.  The heat capacities
of the even-mass $^{58}$Fe, $^{60}$Fe, and $^{62}$Fe all display
 a peak around $T\sim 0.7$ MeV,  which becomes more pronounced
 with an increasing number of neutrons. 

In conclusion, we have introduced a new method for calculating 
the heat capacity in which the statistical errors are strongly
 reduced.  A systematic 
study in several iron isotopes reveals signatures of the pairing 
transition in the heat capacity of finite nuclei despite the 
large fluctuations. 

This work was supported in part by the Department of
Energy grants No.\ DE-FG-0291-ER-40608.
Computational cycles
were provided  by the San Diego
Supercomputer Center (using NPACI resources), and by
the NERSC high performance computing facility at LBL.

\end{document}